\newcommand\anon[2]{{#2}} %known

\documentclass[twoside,leqno,twocolumn]{article}

\usepackage[letterpaper]{geometry}

\usepackage{template_2025/ltexpprt}
\usepackage{hyperref}

\usepackage{cleveref}

\usepackage{algpseudocode} % with algorithmicx, or by itself

\newcommand*\samethanks[1][\value{footnote}]{\footnotemark[#1]}

\usepackage{xcolor}

\usepackage{xspace}
\newcommand{\fclaw}{\texttt{ForestClaw}\xspace}
\newcommand{\pforest}{\texttt{\upshape{p4est}}\xspace}

\usepackage{subcaption}

\usepackage{graphicx}
\graphicspath {{figures/}}
\begin{document}

	\newcommand\relatedversion{}
	\renewcommand\relatedversion{\thanks{The full version of the paper can be
	accessed at
	\anon{\protect\url{redacted-url}}{\protect\url{https://arxiv.org/abs/1902.09310}}}}
	 % Replace URL with link to full paper or comment out this line

	\title{\Large Scalable Mesh Coupling for Atmospheric Wave Simulation}
	\anon{
		\author{Submission ID XYZ}  % replace with actual submission id. Keep
		%things anonymous during the double-blind review.
	}
	{ % Actual author names go below
		\author{Hannes Brandt\thanks{Rheinische Friedrich-Wilhelms-Universit\"at
		Bonn, Germany.}
		\and Tim Griesbach\samethanks
	  \and Matthew Zettergren\thanks{Embry-Riddle Aeronautical University.}
  	\and Scott Aiton\thanks{Boise State University.}
  	\and Jonathan Snively\samethanks[2]
  	\and Donna Calhoun\samethanks[3]
  	\and Carsten Burstedde\samethanks[1]}
	}

	\date{}

	\maketitle

	% Copyright Statement
	% When submitting your final paper to a SIAM proceedings, it is requested
	%that you include
	% the appropriate copyright in the footer of the paper.  The copyright added
	%should be
	% consistent with the copyright selected on the copyright form submitted with
	%the paper.
	% Please note that "20XX" should be changed to the year of the meeting.

	% Default Copyright Statement
	\fancyfoot[R]{\scriptsize{Copyright \textcopyright\ 2025 by SIAM\\
			Unauthorized reproduction of this article is prohibited}}

	% Depending on which copyright you agree to when you sign the copyright form,
	%the copyright
	% can be changed to one of the following after commenting out the default
	%copyright statement
	% above.

	%\fancyfoot[R]{\scriptsize{Copyright \textcopyright\ 20XX\\
			%Copyright for this paper is retained by authors}}

	%\fancyfoot[R]{\scriptsize{Copyright \textcopyright\ 20XX\\
			%Copyright retained by principal author's organization}}

	%\pagenumbering{arabic}
	%\setcounter{page}{1}%Leave this line commented out.

	\begin{abstract}
        We describe the application of a scalable algorithm for
        interpolating solution data in the overlapping mesh region of two
        solvers.  This feature is essential to obtain a globally consistent
        solution for in-situ coupled atmospheric wave simulation.
        We provide timings and discuss a real-world application run.
	\end{abstract}

	\section{Introduction}
	Coupling multiple solvers operating on different, but overlapping, domains
	into a single system is a challenging task often encountered
 	in multiphysics simulations.
        It is commonly addressed in the
	context of the Chimera grid method~\cite{StegerDoughertyBenek83}, which
	partitions the domain of a fluid dynamics simulation into
	overlapping meshes.

	\subsection{Problem specification.}
	We consider mesh coupling for a specific geophysical application, namely an
	atmospheric wave simulation extending from Earth's surface into space
	-- from the troposphere to the thermosphere (and overlying ionosphere).
    The domains of the lower-atmosphere solver MAGIC~\cite{Snively13} and the
    ionosphere solver GEMINI~\cite{ZettergrenSnively15} overlap.
	For a globally consistent solution, they need to exchange interpolated data
	in their
	overlap region, which becomes a significant challenge for
	distributed-memory meshes.
	We use a scalable, forest-of-octrees-based parallel data-exchange
	algorithm to speed up the coupling routine, which was previously performed
	via I/O~\cite{ZettergrenHirschSnivelyEtAl24}.

  \section{Geophysical Simulation Framework}
    \begin{figure}
      \centering
      \includegraphics[width=\columnwidth]{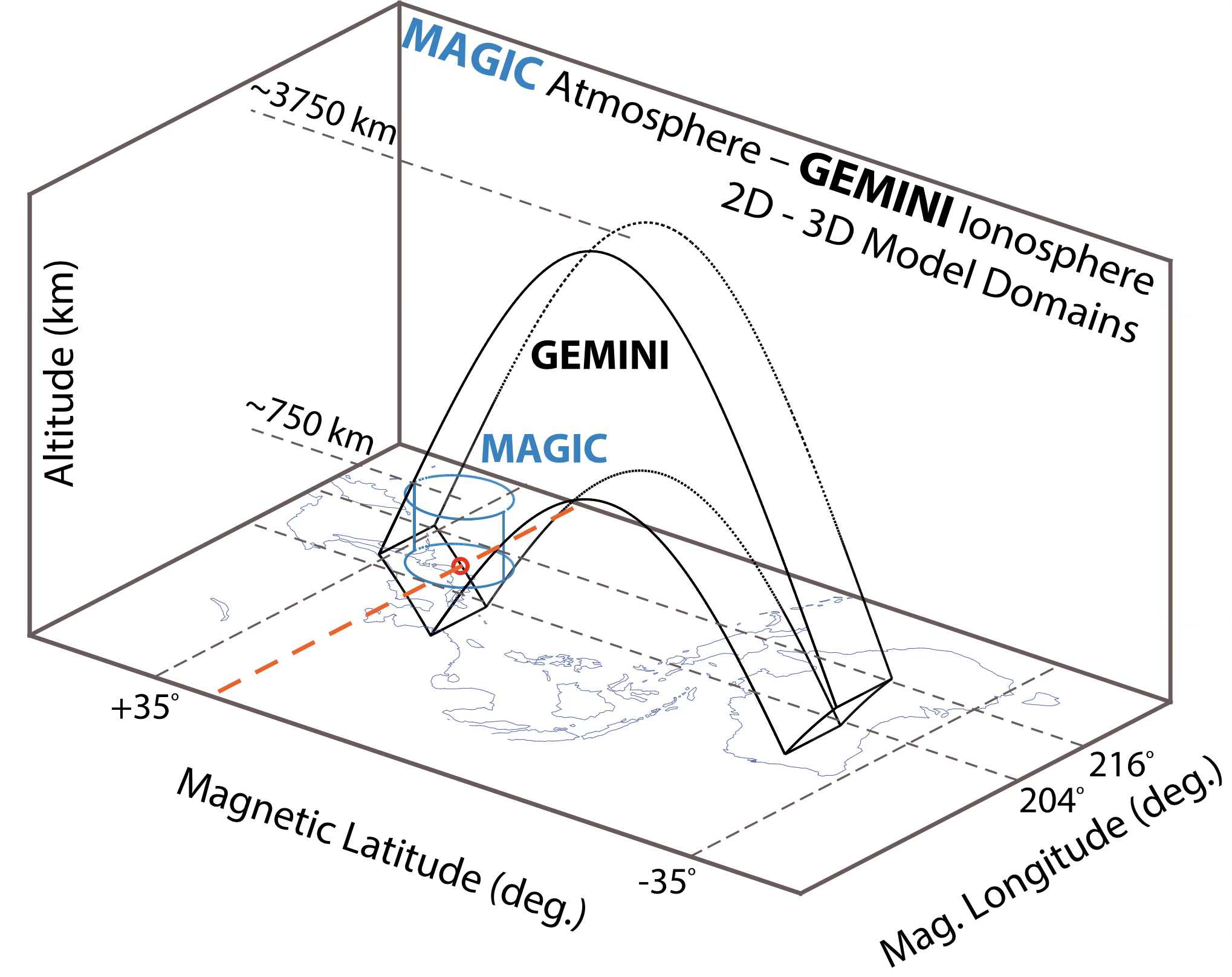}
      \caption{Typical MAGIC and GEMINI domains.}
    \end{figure}

    We consider the  Model for Acoustic-Gravity wave Interactions and Coupling
    (MAGIC) for the simulation of the behavior of
    acoustic gravity waves (AGWs) from their origin that lies in the troposphere,
    i.e.\ from Earth's lowest atmospheric layer, to the thermosphere, and
    extending to the base of the exosphere ($\sim$4000\,km altitude).
    The AGWs are observed via the fluctuations they impose in the ionosphere,
    which overlaps the upper mesosphere and thermosphere. There we employ
    the Geospace Environment Model for Ion-Neutral Interactions (GEMINI)
    to simulate them.

	  \subsection{MAGIC}
    is the model for AGWs generated by transient forcing.
    The waves may be generated by natural hazards, e.g., by earthquakes,
    tsunamis, volcanic eruptions, and explosions, as well as meteorological
    waves~\cite{SnivelyCalhounAitonEtAl23}.
    MAGIC solves the compressible, conservative Navier-Stokes equations for
    nonlinear acoustics from Earth's surface to
    exobase~\cite{SnivelyCalhounAitonEtAl24a}
    and uses a local tangent plane Cartesian coordinate system.

    \subsection{GEMINI}
    is a general-purpose ionospheric model designed to describe medium- and
    small-scale phenomena~\cite{Zettergren25}.
    It uses general orthogonal curvilinear coordinates in either two or three
    dimensions.
    GEMINI couples equations describing the dynamics of ionospheric plasma with
    an electrostatic treatment of auroral and neutral dynamo
    currents~\cite{Zettergren25}.

    \subsection{Software stack.}
    \begin{figure}
      \centering
      % TODO: Add MAGIC and GEMINI to the stack, remove file I/O and checkp.?
      \includegraphics[width=0.89\columnwidth]{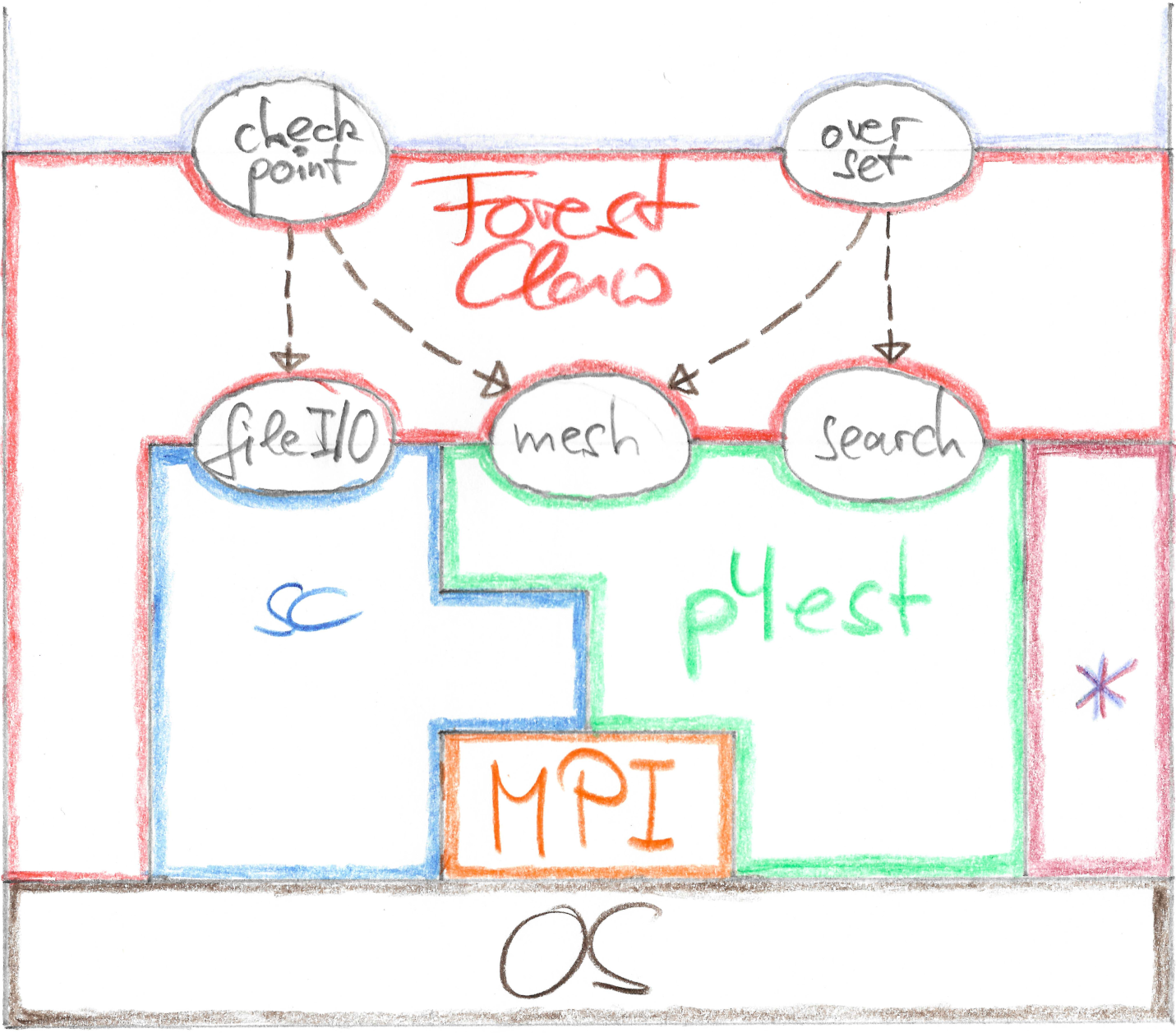}
      \caption{Software stack of \fclaw.}
    \end{figure}
    \fclaw~\cite{CalhounBurstedde17} is a library that combines the
    highly scalable adaptive mesh refinement capabilities (AMR) of the forest-of
    octrees library \pforest~\cite{BursteddeWilcoxGhattas11} and the clawpack
    solver~\cite{Clawpack25}.

    In the MAGIC Forest~\cite{SnivelyCalhoun21} library, MAGIC
    is implemented in \fclaw, thereby adding scalable AMR
    capabilities to its functionality and achieving significant speedups.
    Additionally, the adaptation to \fclaw enables 3D extruded and full 3D
    versions of MAGIC.

    In Trees GEMINI~\cite{ZettergrenHirschSnivelyEtAl24}, the GEMINI library is
    combined with \fclaw, so it can use its mesh refinement capabilities for
    static
    refinement around known locations of interest and adaptive refinement for
    evolving fields. Similar to MAGIC Forest, it can create 3D extruded and
    full 3D as well as 2D meshes.

	\section{Coupling of MAGIC and GEMINI}
    To combine MAGIC and GEMINI in a single system, they need to exchange
    simulation data in their region of overlap to ensure a consistent solution.
    Since both solvers rely on mutually unrelated parallel mesh partitions,
    executing the exchange is a hard problem at scale.
    This was previously dealt with using a file I/O routine and subsequent
    root-worker data distribution.
    The switch to \fclaw-based MAGIC and GEMINI allows us to access an
    algorithm we designed for the exchange of data between two parallel
    meshes and achieve a significant speedup of the process.

    \subsection{Exchange algorithm.}
    \begin{figure}
      \centering
      \includegraphics[width=\columnwidth]{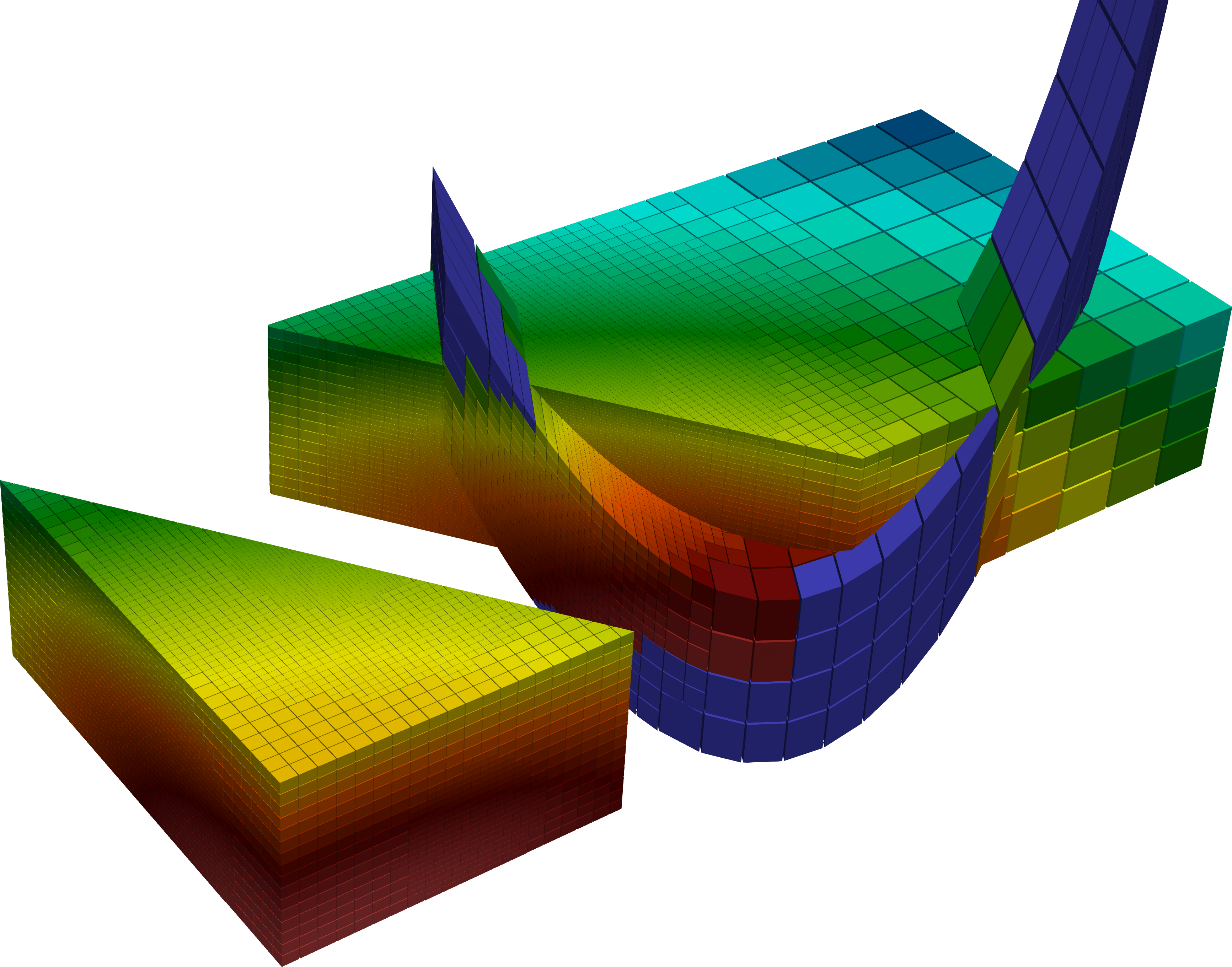}
      \caption{Exchange between two forests of octrees.}%with color-coded data.}
    \end{figure}
    The \fclaw library employs a patch-based AMR approach implemented on top
    of the forest-of-octrees AMR structure of
    \pforest~\cite{CalhounBurstedde17}.
    Every leaf of the forest corresponds to one patch, each a uniform grid
    of cells.

    We implement a modular exchange algorithm~\cite{BrandtBurstedde26}
    operating on a forest and
    another mesh of arbitrary structure.
    The forest of octrees provides data to the other mesh.
    Accordingly, they are referred to as producer and consumer, respectively.
    The consumer mesh creates distributed sets of query points, e.g., by
    generating one query for each cell in a patch.
    Next, the queries are searched in the distributed forest of octrees
    mesh.
    For this, we leverage an efficient, communication-free partition search
    routine of \pforest~\cite{Burstedde20d} that allows to
    assign each query to the process that owns the patch containing it.
    Based on these results, the queries are sent to the corresponding process
    and
    searched in its local partition \cite{IsaacBursteddeWilcoxEtAl15}.
    Whenever a query is found in a leaf patch of the forest, it is supplied with
    data.
    Subsequently, all queries are returned together with the received data to
    their process of origin using non-blocking point-to-point communication.

    The algorithm uses an opaque point structure for maximal
    flexibility.
    Accordingly, the parallel search routine relies on a user-defined
    intersection callback that determines, for a given query and a given patch,
    whether these intersect.
    The callback has to take the application-dependent coordinate systems into
    account.
    If required, the callback has to first convert the query point and the patch
    to the same coordinate system.
    Since the query commonly represents a single point in space, it is often
    easier
    to convert the point into the reference domain of the patch than the other
    way around.
    Similar to the intersection tests, the interpolation on leaf level is also
    left to a user-defined interpolation callback.

    \subsection{Application to MAGIC and GEMINI.}
    \begin{figure*}[ht]
	    \begin{minipage}[t]{.4\textwidth}
		    \centering
		    \includegraphics[width=\textwidth]{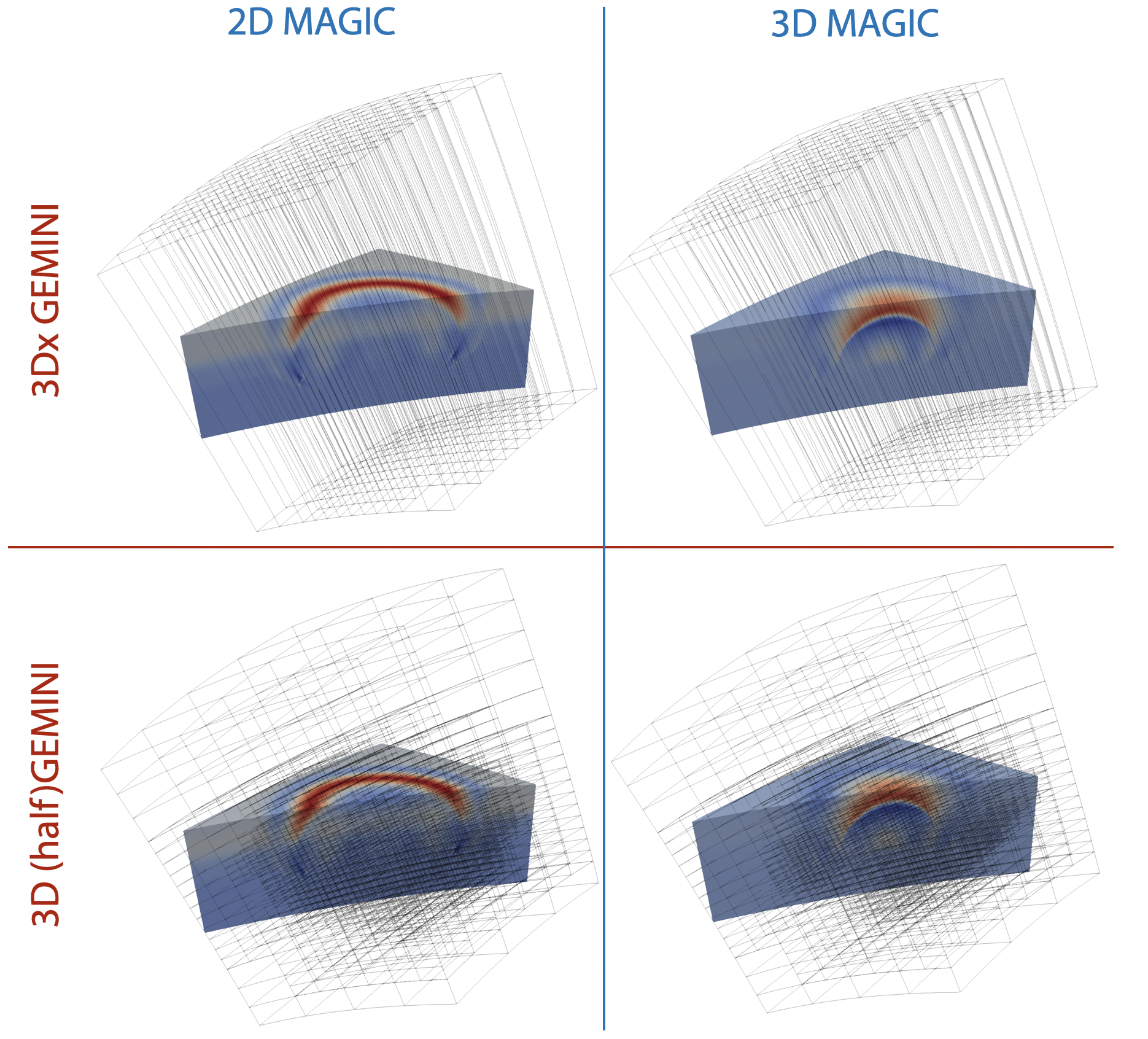}
		    \caption{Combining coupling dimensions.}
		    \label{fig:overset_modes}
		  \end{minipage}
		  \begin{minipage}[t]{.6\textwidth}
		    \centering
		    \includegraphics[width=0.9\textwidth]{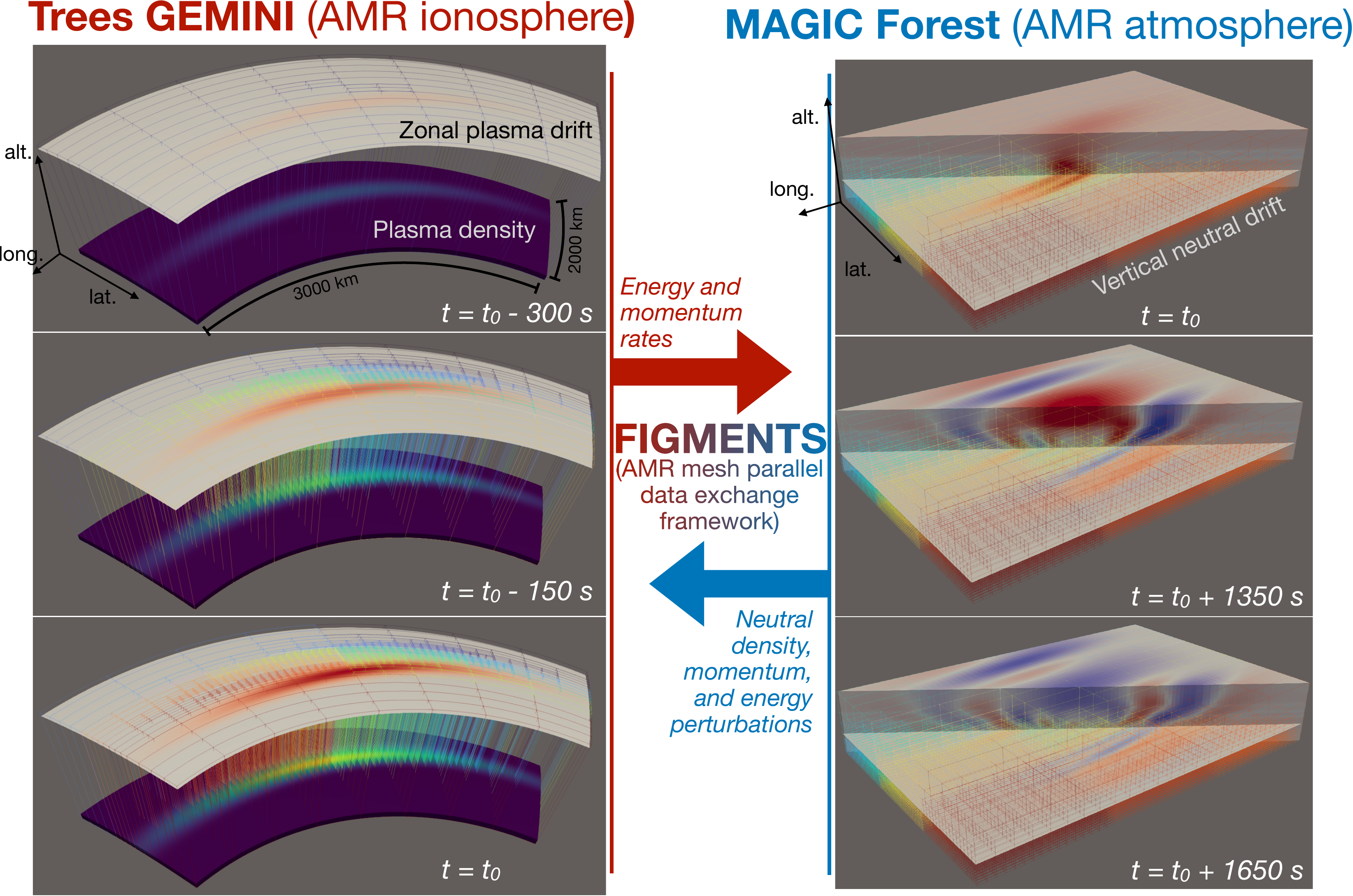}
		    \caption{Two-way mesh data exchange.}
		    \label{fig:two_way}
		  \end{minipage}
    \end{figure*}

    To combine MAGIC and GEMINI in a single simulation setup, GEMINI has to
    receive data interpolated from MAGIC's solution, i.e.\ neutral
    density, momentum and energy perturbations.
    We can use the \fclaw implementation \cite{Calhoun25}
    of the exchange algorithm for this task.
    Here, GEMINI takes the role of the consumer, while MAGIC is the producer.

    In a first step, we iterate over all memory-local patches of the GEMINI
    mesh and
    query the center of every cell in the patch.
    The queries are directly converted from the generalized orthogonal
    coordinate system employed by GEMINI into the East North up (ENU) local
    tangent plane Cartesian coordinate system used by MAGIC.
    In the case of a curvilinear dipole coordinate
    system~\cite{HubaJoyceFedder00}, as commonly used in GEMINI simulations,
    this transformation is done by converting to Spherical Earth-centered,
    Earth-fixed (ECEF) coordinates and then to Cartesian ECEF coordinates as
    intermediate steps.

    Next, the queries are entered into the parallel search routine.
    To identify matches between the MAGIC Forest patches and the query points
    created by GEMINI, we have to provide an intersection callback.
    First, the callback converts the query points given in MAGIC's physical
    coordinate system into the reference coordinate system of the tree to which
    the
    given patch belongs.
    As MAGIC uses local Cartesian coordinates that are mapped to the unit
    square, this inversion process equals a simple scale-and-shift operation
    based on the tree of the forest we are in.
    Next, the actual intersection test is performed for the mapped query point
    and the axis-aligned patch given in reference coordinates.
    Whenever a query point matches a patch on leaf level, trilinear
    interpolation from the patch cells to the query points is performed for all
    required data.

		The modular implementation of the exchange allows to search the same set of
		GEMINI queries in a 2D, 3D or 3D extruded MAGIC domain (see
		Figure~\ref{fig:overset_modes}) with only slight
		adaptations to the intersection and interpolation callbacks.
		Switching between a 3D and a 3D extruded GEMINI domain is even simpler, as
		this change only affects the creation of query points.

		Although viable, the exchange does not need to be called after every
		time step
		of the two solvers.
		Instead, both solvers can compute several iterations of their individual
		time-stepping scheme, until they both exceed a predetermined time
		threshold.
		Then a single exchange is performed, before the procedure starts again.

    \subsubsection{Two-way exchange.}

    As Trees GEMINI is based on \fclaw similar to MAGIC Forest, it is possible
    to exchange information between the meshes in both directions.
    For this, we follow up on the previously described exchange routine with a
    second call employing swapped roles.
    Here, MAGIC creates a query for every cell center and converts it to
    GEMINI's orthogonal curvilinear coordinate system.
    During the exchange, the queries are provided with interpolated
    energy and momentum rates from GEMINI; see Figure~\ref{fig:two_way}.

  \section{Results}
    We perform a 2D AGW simulation on 384 cores.
    MAGIC is initialized for our timing test with an acoustic pulse, specified
    as an axisymmetric initial condition about the origin, to represent the
    wave field of an explosive source (other sources used with the models
    include convective plumes or earthquake surface motions). The MAGIC
    atmosphere is specified empirically in a domain 0-400\,km in radius and
    altitude; the overlying GEMINI model is initialized to a steady-state that
    evolves self-consistently from the ambient neutral
    state~\cite{SnivelyCalhounAitonEtAl25,ZettergrenCalhounSnivelyEtAl25}.
    The waves from the initial pulse are tracked adaptively as they propagate
    through MAGIC’s mesh, over 10 levels of refinement ($<$\,4\,m to 4\,km
    resolution), reaching the GEMINI ionosphere around $t=300$\,s. There,
    GEMINI’s mesh refines to track the MAGIC wave field with $\sim$\,10\,m
    resolution, as it expands and also dissipates. The evolving transient
    acoustic pulse provides a feasible scalability test, as it spans a wide
    range of scales and number of patches during a simulation.
    The total number of MAGIC patches ranges
    from 33,931 to 63,064 and the number of GEMINI patches ranges from 4,376 to
    20,459.
    The number of cells, and thus queries, per GEMINI patch is constant at 128.
    \begin{figure}
      \centering
      \includegraphics[width=\columnwidth]{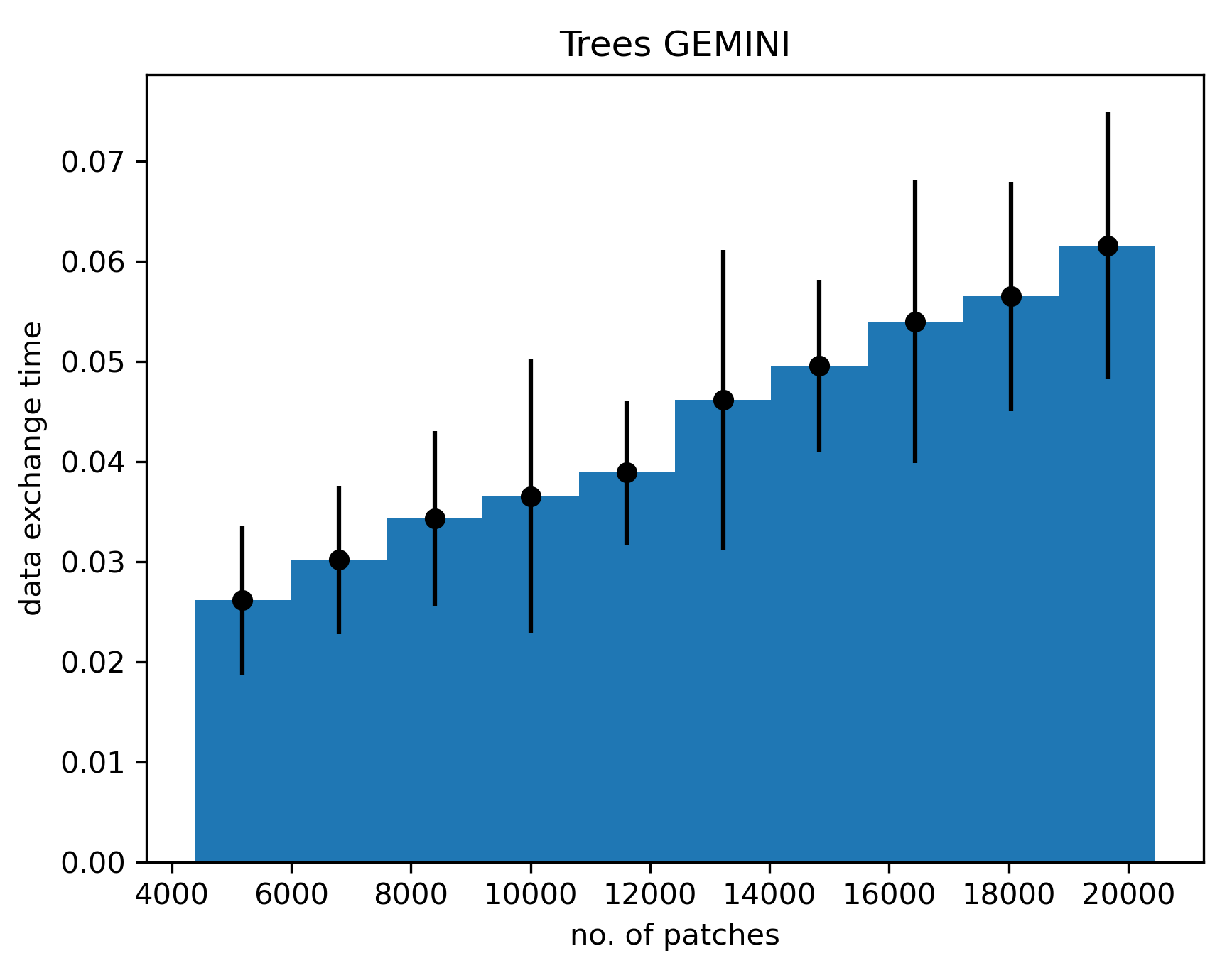}
      \caption{Exchange run times in wall clock seconds for varying global
               GEMINI patch counts on 384 processes.}
      \label{fig:scaling}
    \end{figure}

    The workload of the exchange depends heavily on the number of queries that
    have to be searched and supplied with data.
    This matches the linear dependence between the number of GEMINI patches and
    the average exchange run time that can be observed in
    Figure~\ref{fig:scaling}.
    The exchange time spent per GEMINI patch even decreases for higher total
    patch counts,
    which matches the eventually decreasing portions of patches in the overlap
    region due to the center of refinement shifting away.
    Additionally, load imbalances of the exchange and the overall low patch
    count per process need to be taken into account.
    In synthetic tests of its \pforest implementation, the exchange algorithm
    showed scaling to 12,288 memory-distributed
    workers~\cite{BrandtBurstedde26}.
    It was applied to both 2D and 3D test cases.
    Moreover, the scalability improved significantly after weighted
    repartitioning of the forest to account for problem-inherent load
    imbalances.

    Next to the algorithm's scalability the overall run time is of interest as
    well.
    The exchange took a total of $0.04$\,s seconds on average, while an average
    MAGIC time step required $0.83$\,s and an average GEMINI time step required
    $0.71$\,s.
    The impact of the exchange on the total run time can be reduced even
    further, as it is not necessary to synchronize the two solvers after every
    time step.
    Throughout the simulation, there were on average 19 MAGIC and 27 GEMINI
    time steps in between synchronizations.

  \section{Conclusion}
    We apply a scalable exchange algorithm, employing a forest of
    octrees, to mesh coupling of the \fclaw-based versions of the MAGIC and
    GEMINI libraries.
    The resulting method is able to interpolate data from both meshes and
    handle
    meshes of different dimensionalities.
    In an AGW simulation,
    it showcases good scalability and a negligible run time overhead.

  \section*{Acknowledgements}
    The authors gratefully acknowledge partial support under DARPA Cooperative
    Agreement HR00112120003 via a subcontract with Embry-Riddle Aeronautical
    University.
    This work is approved for public release; distribution is unlimited.
    The information in this document does not necessarily reflect the position
    or the policy of the US Government.

    We acknowledge additional funding by the Bonn International Graduate School
    for Mathematics~(BIGS) as a part of the Hausdorff Center for
    Mathematics~(HCM)	at the University of Bonn.
    The HCM is funded by the German Research Foundation~(DFG) under Germany’s
    excellence initiative EXC 59 – 241002279 (``Mathematics: Foundations,
    Models, Applications'').
    We acknowledge further funding by the DFG under grant no.\ 467255783
    (``Hybride AMR-Simulationen'').

    We would also like to acknowledge access to the ``Bonna'' and ``Marvin''
    compute clusters hosted by the University of Bonn.%

  \bibliographystyle{siam}
  \bibliography{group,ccgo_new}

\end{document}